\documentclass[preprint]{aastex}

\newcounter{species}

\usepackage{hyperref, enumerate}
\hypersetup{colorlinks,linkcolor=blue, citecolor=blue}

\usepackage{url}
\usepackage{color}
\usepackage{epstopdf}

\shorttitle{Modal Noise Solution }
\shortauthors{Mahadevan et al.}
\begin{document}
\title{Suppression of Fiber Modal Noise Induced Radial Velocity Errors for Bright Emission-Line Calibration Sources}

\author{Suvrath Mahadevan\altaffilmark{1,2,3},
Samuel Halverson \altaffilmark{1,2,3},
Lawrence Ramsey\altaffilmark{1,2}, 
Nick Venditti \altaffilmark{1}
}

\email{suvrath@astro.psu.edu}
\altaffiltext{1}{Department of Astronomy and Astrophysics, The Pennsylvania State University, 525 Davey Laboratory, University Park, PA 16802, USA.}
\altaffiltext{2}{Center for Exoplanets and Habitable Worlds, The Pennsylvania State University, University Park, PA 16802, USA.}
\altaffiltext{3}{Penn State Astrobiology Research Center, The Pennsylvania State University, University Park, PA 16802, USA. }

\begin{abstract}

Modal noise in optical fibers imposes limits on the signal to noise and velocity precision achievable with the next generation of astronomical spectrographs. This is an increasingly pressing problem for precision radial velocity (RV) spectrographs in the near-infrared (NIR) and optical that require both high stability of the observed line profiles and high signal to noise. Many of these spectrographs plan to use highly coherent emission line calibration sources like laser frequency combs and Fabry-Perot etalons to achieve precision sufficient to detect terrestrial mass planets. These high precision calibration sources often use single mode fibers or highly coherent sources. Coupling light from single mode fibers to multi-mode fibers leads to only a very low number of modes being excited, thereby exacerbating the modal noise measured by the spectrograph.  We present a commercial off-the-shelf (COTS) solution that significantly mitigates modal noise at all optical and NIR wavelengths, and which can be applied to spectrograph calibration systems.  Our solution uses an integrating sphere in conjunction with a diffuser that is moved rapidly using electrostrictive polymers,  and is generally superior to most tested forms of mechanical fiber agitation.  We demonstrate a high level of modal noise reduction with a narrow bandwidth 1550 nm laser. Our relatively inexpensive solution immediately enables spectrographs to take advantage of the innate precision of bright state-of-the art calibration sources by removing a major source of systematic noise.

\end{abstract}

\section{Introduction}
\label{introsection}
 High-resolution fiber fed spectrographs are being used for demanding astrophysical applications, including the detection of low mass planets (eg. \citealt{dumusque12}) using precise radial velocity (RV) measurements. The promise of detecting rocky planets around M dwarfs has led to the development of stabilized high resolution near-infrared (NIR) spectrographs such as the Habitable Zone Planet Finder (HPF, \citealt{mahadevan12}) and CARMENES \citep{quirrenbach12}, while ambitious instruments like ESPRESSO \citep{megevand12} seek to achieve 10 cm s$^{-1}$ RV precision to  find true Earth-analogs around G and K stars. These instruments require extremely stable and accurate calibration sources like laser frequency combs and stabilized Fabry Perot etalons that are now being developed and tested.
 
  The finite number of electromagnetic modes in an optical fiber leads to a form of noise that, if left unmitigated,  limits the achievable signal to noise and adds a source of RV noise. This {\it modal noise} is worst with narrow emission line sources, and at longer wavelengths, and can significantly hinder the ability to achieve the high signal to noise needed to search for biomarkers in the atmosphere of transiting planets \citep{snellen13} in the NIR.  
  
  In this paper we experimentally demonstrate a commercially available and easy to implement solution to modal noise that is applicable to calibration sources. This solution is immediately useful to multiple groups attempting to push forward the limits of  precision spectroscopy in the optical and NIR. The development of such techniques is closely related to the growing field of {\it Astrophotonics} \citep{2009OExpr..17.1880B}.

\section{Fiber Modal Noise \& its Impact on High Precision Spectroscopy }
\label{modalnoisesection}

The term {\it modal noise} refers to a noise source caused by  the changing spatial distribution of light at the output end of a fiber. The conditions under which modal noise manifests itself \citep{rawson80} are: 
\begin{itemize}
\item A narrow source or observed spectrum.
\item Spatial filtering of the light from the output face of the fiber.
\item Movement, stress or temperature changes in the fiber, wavelength change in the source, or changes in input illumination that change the distribution of propagating modes.
\end{itemize}

All three conditions are present in modern fiber-fed astronomical spectrographs to some degree, and lead to a manifestation of modal noise in the recorded spectra \citep{baud01, grupp03}. Calibration light input to the spectrograph can be highly coherent and narrow-line width (eg. laser frequency combs). Some spatial filtering is always present in any spectrograph, typically caused either by a slit, spatially dependant grating efficiency, vignetting, variable path-length through prisms, or other similar sources.  Optical fibers have to connect the spectrograph to the telescope focal plane and will inevitably move and bend as the telescope slews. Temperature changes and stress in the fibers will also cause the output speckle pattern to change, all resulting in effectively limiting the achievable signal to noise of the observed spectrum.

While the term modal noise traditionally refers to fluctuations in the recorded intensity, the presence of a finite number of speckles is also a source of noise in determining the centroid of an observed emission and absorption line, even with no spatial filtering. Fiber-fed astronomical spectrographs dedicated to high precision RV surveys are affected by both the limitation in signal to noise and the RV noise.

We distinguish modal noise from {\it \lq fiber scrambling\rq}  as the term is commonly used in astronomical spectroscopy. The former arises from the result of the finite number of electromagnetic modes excited in optical fibers, and the conditions described above, while the latter refers to the dampening of correlated spatial variations in fiber output when the input illumination varies. Optical double scramblers \citep{hr92} that interchange the near-field and far-field images between two fibers to provide both radial and azimuthal scrambling, are now routinely used for the highest precision spectrographs. Coupling these with non-circular fibers (eg. octagonal fibers) provides additional radial and azimuthal scrambling leading to output illumination that is largely decoupled from the input variations, enabling high velocity precision \citep{bouchy13}. These techniques {\bf do not mitigate modal noise} since such scrambling is static and merely redistributes the modes.  \citet{mccoy12} have shown that the use of octagonal fibers does not provide any better modal noise suppression than the use of circular fibers.

 Following \citet{gr81} and \citet{lemke11},  the modal noise limited signal to noise ($SNR$)  from a fiber is

\begin{equation}
{\mathrm {SNR}} = \frac{\rho}{\nu} \sqrt{\frac{M+1}{1-\rho^2}}
\end{equation}

where $\rho^2$ is the fraction of light from the illuminated fiber end incident on the detector, $M$ is the number of excited modes in the fiber, and $\nu$ the visibility of the speckle contrast at the fiber output. For highly coherent narrow linewidth calibration sources it is appropriate to set $\nu \sim 1$. For illumination of high resolution spectrographs by less coherent sources (ie. starlight), the expected speckle contrast (or visibility) is lower, leading to modal noise being less of a problem, though not negligible. For such sources \citet{lemke11} experimentally derive a visibility ($\nu$) ranging from 0.01-0.05.   The  $\mathrm{SNR}$ approaches infinity as $\rho^2$ approaches 1, implying that the intensity modal noise is no longer the limiting factor compared to other noise sources like the intrinsic photon noise. In reality  $\rho^2$ is never unity for any astronomical spectrograph. 

The maximum number of modes (for unpolarized light) at a wavelength $\lambda$ supported by a cylindrical step index fiber of diameter $d$, and input numerical aperture $\mathrm NA$ is given by 

\begin{equation}
M=0.5 \left(\frac{\pi d {\mathrm NA}}{\lambda}\right)^2
\end{equation}

\begin{figure}
\begin{center}
\includegraphics[scale=.6]{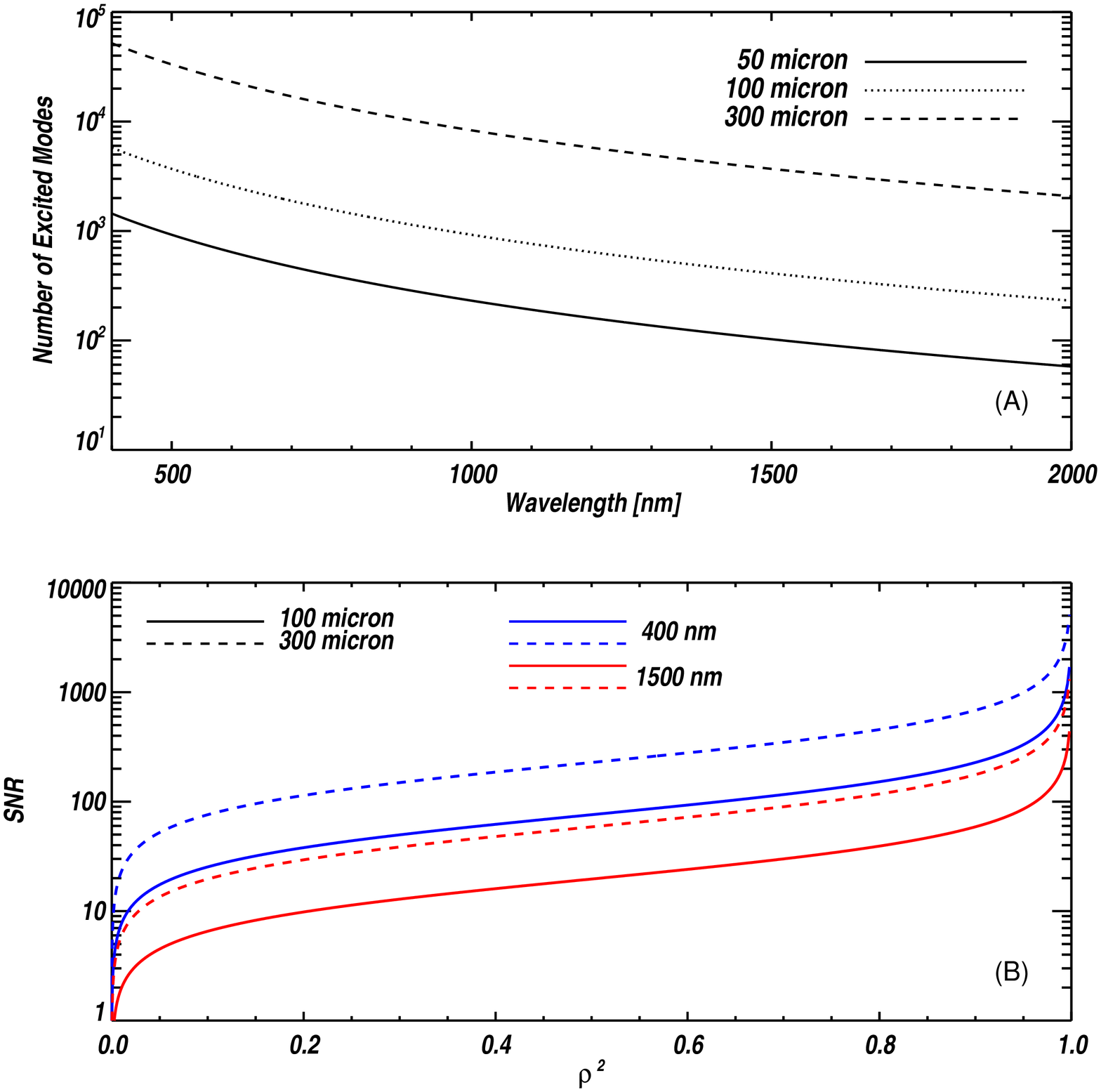}
\caption{(A) Number of modes propagating at each wavelength in step index cylindrical optical fibers of diameter 50 $\mu$m, 100 $\mu$m and 300 $\mu$m for unpolarized light. For simplicity we have assumed an input focal ratio of f/3.65 (NA=0.1357) in all cases (B) The maximum achievable signal to noise with a narrow linewidth source for 100 $\mu$m and 300 $\mu$m fibers at wavelengths of 400 nm and 1550 nm, chosen to roughly correspond to the limits of wavelengths covered by precision RV spectrographs. $\rho^2$ is the fraction of light from the illuminated fiber end incident on the detector\label{modesinfiber}}
\end{center}
\end{figure}

Figure \ref{modesinfiber} shows the number of modes propagating at each wavelength in a 300$\mu$m (HPF), 100$\mu$m (CARMENES), and 50$\mu$m  fiber at an input focal ratio of f/3.65. Most high resolution instruments operate at input focal ratios between f/3.5 and f/5 to avoid significant amounts of focal ratio degradation. Also shown is the corresponding SNR for coherent input illumination (such as that provided by a laser frequency comb),  as a function of $\rho^2$ for wavelengths of 400 nm and 1550 nm. As expected, the modal noise is significantly higher for the NIR wavelengths due to the much smaller number of available modes ($M$) in the fiber. Using a slit or an image slicer leads to a lower $\rho^2$, thereby increasing modal noise on the recorded spectra.  Noise increases even in the case of an ideal (zero-loss) image slicer with an ideal spectrograph. The speckle distribution on the fiber output is constrained by the need to preserve the total intensity \citep{gr81}, but no such constraint applies to any of the individual fiber slices. 

\section{Wavelength Calibration Sources}
\label{wavecalsection}
Challenging astrophysical goals push high-resolution spectrographs to use the most precise wavelength calibration sources available. 
For fiber-fed spectrographs the calibration source sets the wavelength scale and helps continuously monitor the instrument drift. Commonly used atomic emission line lamps like Thorium-Argon and  Uranium-Neon \citep{redman11} provide a large number of stable emission lines across the UV-NIR range. 
  
Laser frequency combs offer significant calibration advantages in astronomical spectroscopy \citep{murphy07, braje08}.  Such combs are now being built and tested both with optical \citep{locurto12} and NIR \citep{osterman12} spectrographs, and can lead to significant  improvements in RV precision. The intrinsic linewidths of each of these laser comb lines is typically $\sim$300-800 kHz \citep{quinlan10, ycas12b} making them extremely coherent narrow band sources, and therefore subject to significant modal noise. An added complication with these devices is that the light is output in a single-mode fiber (SMF hereafter). Coupling of the SMF to a multi-mode fiber excites only a small number of modes in the multi-mode fiber, further exacerbating the impact of modal noise.   The problem can be mitigated by agitating the optical fiber \citep{baud01}. While this has been done at the shorter optical wavelengths, with good results, systematic noise and velocity error has been reported at longer wavelengths \citep{phillips12a,  phillips12b}. Tests with the Pathfinder spectrograph and a laser comb in the NIR H band (1.5--1.7 $\mu$m) were limited by modal noise even with the use of an integrating sphere and a high frequency fiber agitator \citep{ycas12, redman12}. 

Many groups are also pursuing the development of calibration sources based on Fabry-Perot etalons \citep{wildi12, halverson12}. While lacking the level of stability and frequency traceability of a comb, these sources can be stabilized at a high level, at least short term, enabling the spectrograph drift to be monitored at high precision.  The linewidths from these devices  range from 0.1-1GHz; still  less than the resolution element of the spectrographs. The speckle visibility $\nu$ is still very high, and modal noise remains a significant problem.

\section{Dynamic Modal Noise Mitigation}

\citet{mccoy12} have discussed the mitigation of modal noise by fiber agitation.  In the NIR bulk agitation of fibers is necessary, even with the use of an integrating sphere, and a specific form of high amplitude low frequency hand agitation performs better than currently tested forms of mechanical agitation. While more sophisticated mechanical agitators will eventually solve this problem,  these devices are harsh on the optical fibers and may  reduce their lifetime. In addition a source of high level of vibrations near, or coupled to, a  stabilized spectrograph is undesirable. This lead us to explore  dynamic diffusers that rapidly change the input modal pattern, thereby effectively randomizing the output mode pattern.

Our experiments achieve a high level of modal noise mitigation with a commercially available dynamical diffuser, which is an Optotune\footnote{http://www.optotune.com/} Laser Speckle Reducer \citep{blum12}.  These devices consist of a  diffusive surface attached to electroactive polymers that are used as actuators to rapidly move the diffuser.  Voltage applied across the actuators leads to a squeezing of the elastic polymer film, causing them to expand laterally and move the attached diffuser. This can be done very rapidly ($\sim$ 180--300 Hz), causing rapid changes in the output speckle pattern. Commercially available diffusers have a surface relief structure that is tuned to the required performance, such as angle of diffusion, shape of the diffused beam, etc. In general a larger diffusion angle requires smaller diffuser structures. The changes in the output pattern can further be randomized by the use of an integrating sphere. The key concept here is that while an integrating sphere alone randomizes the output modal distribution, that randomization is static, leaving one still susceptible to modal noise in the fibers.  A combination of a dynamic diffuser {\it \bf and} integrating sphere  yields dramatic reductions in modal noise.

We purchased an Optotune Laser Speckle Reducer with a specified diffusion angle of $20^\circ$ and a 10 mm clear aperture. For our experiment  we used a 1550 nm laser with a specified line width of  $<10$MHz, which is expected to lead to a high speckle contrast ($\nu=1$). Working at 1550 nm also ensures we are probing modal noise at long wavelength, where it is expected to be worse. We use a single mode fiber (SMF28) to couple the laser light through the diffuser and into an  4 inch diameter integrating sphere, and use a Polymicro  $200 \mu$m core fiber on the output fiber port of the integrating sphere. The integrating sphere was purchased from SphereOptics\footnote{http://www.sphereoptics.de}, and is the same sphere used in on-sky demonstrations of a laser frequency comb described in \citet{ycas12}. The sphere is manufactured using a proprietary form of Teflon that approaches ideal Lambertian behavior and provides a diffuse reflectance value of 98\% acoss the visible and infrared.    The output of the fiber is re-imaged onto a Xenics InGaAs NIR camera that records the speckle pattern.  The choice of the 200 $\mu$m fiber was driven by availability as well as the fact that the number of modes propagating in this fiber, when fed at its maximum numerical aperture (NA=0.22) by the integrating sphere, is quite similar to that propagating in a 300 $\mu$m fiber fed at f/3.65 (which is the planned fiber for the HPF spectrograph). 

\begin{figure}
\begin{center}
\includegraphics[scale=.4]{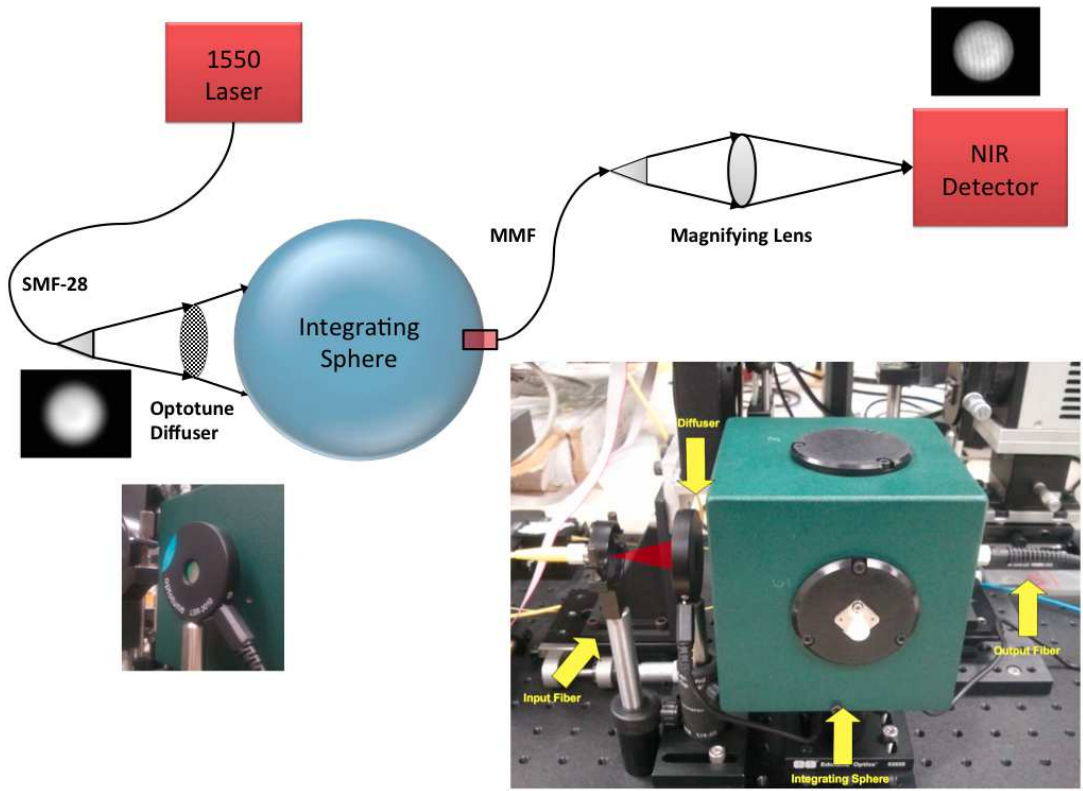}
\caption{Sketch of the experimental layout with image insets showing the diffuser next to the integrating sphere. Light from the laser is coupled by an SMF fiber through the diffuser, into the integrating sphere, and into a multi-mode fiber.  \label{sketch}}
\end{center}
\end{figure}

Figure \ref{sketch} shows a sketch of the experimental setup as well as an inset image of the diffuser. The experimental setup allows use of the diffuser with and without the dynamical motion, removal of the diffuser, and the integration of a mechanical agitator that is an improved version of one we described in \citet{mccoy12}.

\begin{figure}
\begin{center}
\includegraphics[scale=.6]{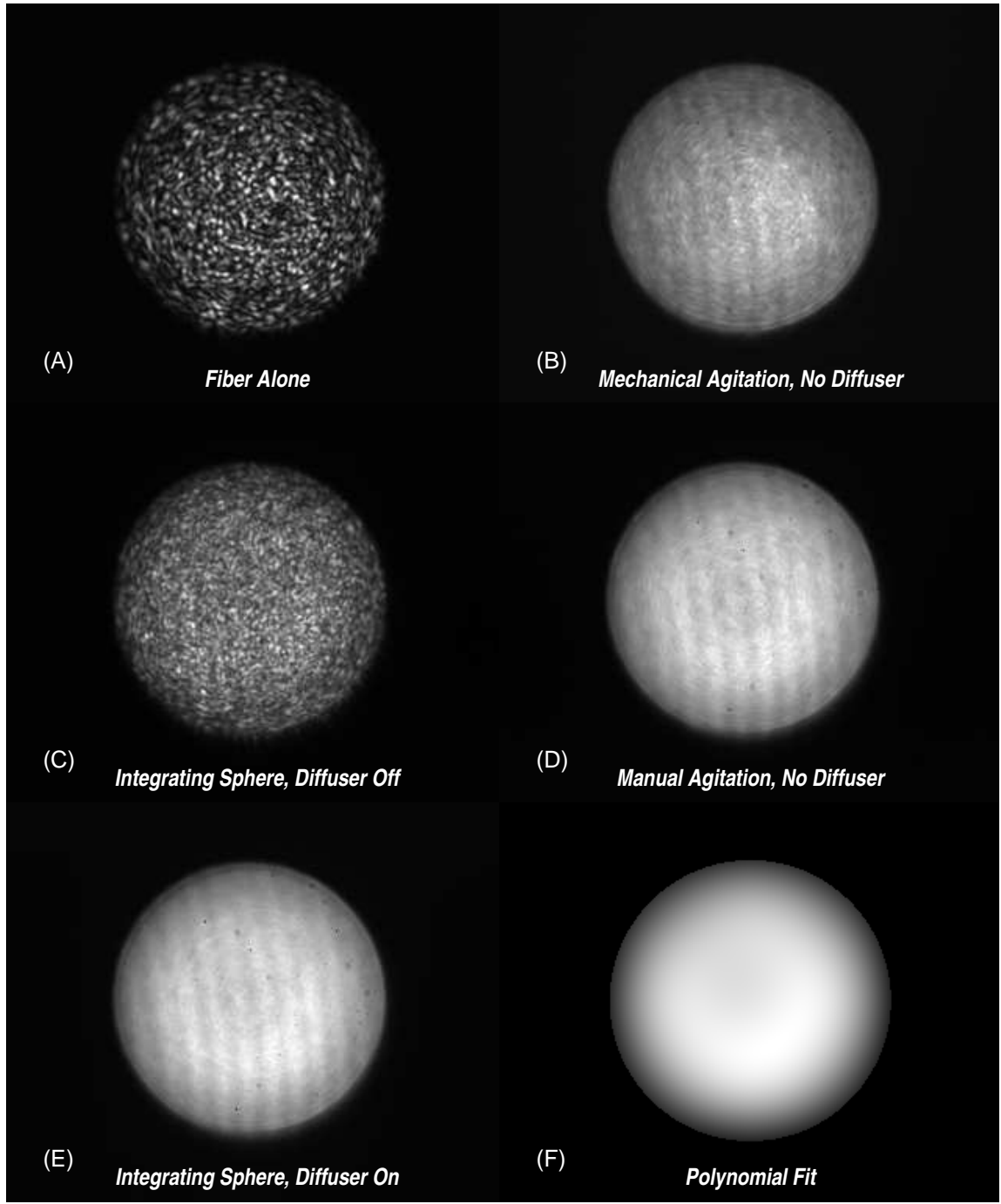}
\caption{Images of the spatial distribution of light for (A) SMF directly coupled to the un-agitated multi-mode fiber (B) SMF coupled directly to the integrating sphere, and mechanical agitation of the multimode fiber (no diffuser) (C) SMF coupled to integrating sphere through the static diffuser (D) SMF coupled directly to the integrating sphere (no diffuser), and hand agitation of the multimode fiber (E) SMF coupled to integrating sphere through the dynamic diffuser (F) A synthetic image constructed using low order polynomials fit to the previous images to determine the achievable floor to modal noise mitigation.  \label{fiberimage}}
\end{center}
\end{figure}

\begin{figure}
\begin{center}
\includegraphics[scale=.7]{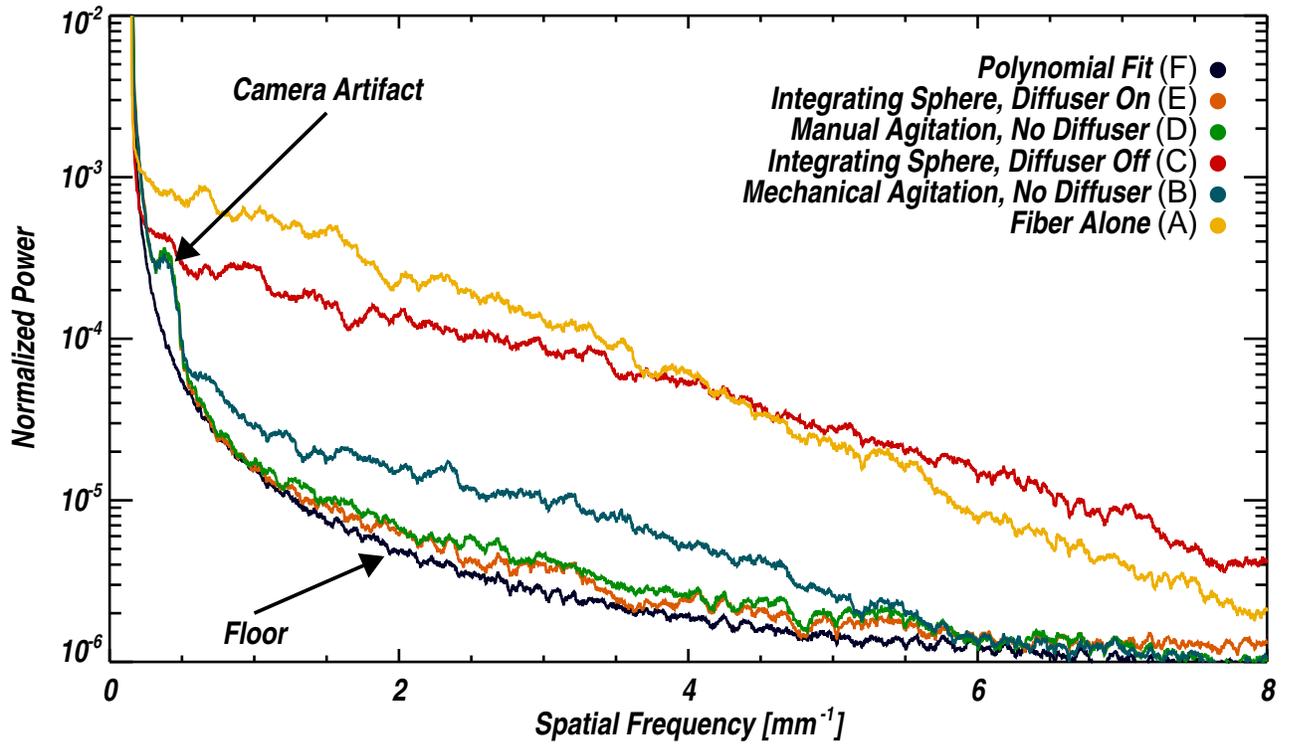}
\caption{Azimuthally averaged power spectrum of images in Figure \ref{fiberimage}. The fit to the power spectrum of the smooth polynomial image fit is shows as a black line and sets the achievable floor to modal noise mitigation.  The static diffuser does not significantly reduce the higher frequencies, whereas the dynamic diffuser coupled with the integrating sphere reduced modal noise to almost within a factor of two of the expected floor. \label{powerspectrum}}
\end{center}
\end{figure}

As part of our experiment we obtained short integration time (3 seconds) images of the modal pattern from the output fiber for five different tests:
\begin{enumerate}[(A)]
\item Direct injection of the light from the SMF fiber to the multimode fiber
\item SMF coupled directly to the integrating sphere, and mechanical agitation of the multimode fiber
\item SMF coupled to integrating sphere through the diffuser, but no diffuser motion (static diffuser) 
\item SMF coupled directly to the integrating sphere, and hand agitation of the multimode fiber
\item SMF coupled to integrating sphere through the diffuser, with diffuser actuated motion (dynamic diffuser)
\end{enumerate}
These images are shown in Figure \ref{fiberimage}. As can be seen visually the dynamic diffuser or manual hand agitation leads to a significant reduction in speckle contrast compared  to other cases. Figure \ref{powerspectrum} plots the azimuthally averaged power spectrum of the  images, showing the significant reduction in the intensity at intermediate and high spatial frequencies.  Figures \ref{fiberimage},\ref{powerspectrum} both show the impact of fringing in the experimental setup, likely from the detector window, which does not impact the results. The theoretical floor is set by a smooth fit to the power spectrum of the synthetic image (Figure \ref{fiberimage}F). The dynamic diffuser performs extremely well, approaching this floor. A specific form of hand agitation at $\sim 1-2$ Hz with a $\sim10$cm fiber bend is also able to approach these levels, but is not highly repeatable and not a practical alternative for a long term RV survey.  Based on these results we are confident that the use of dynamic diffusers effectively removes any systematic based SNR limitation on high resolution spectroscopy even with highly coherent narrow line sources in the NIR H band, corresponding to a worst case manifestation of modal noise. It follows that the problem is also effectively solved for any shorter wavelengths and for less coherent calibration sources. 

The solution can be applied to any calibration source, but with a significant light loss penalty. While the diffuser transmits 60-70\% of the light, the integrating sphere is typically very lossy.  In our experiments the total efficiency of the system is $\sim10^{-6}$, driven largely by the integrating sphere. This is not necessarily a problem. Laser combs and Fabry Perot cavities fed by supercontinuum sources are often very bright, enabling sufficient photons to be recorded by the spectrograph even with the losses from the integrating sphere. 

Typical NIR laser frequency combs now under development can conservatively output $\sim$1-10 nW per individual comb mode. Factoring in losses from the integrating sphere, this roughly corresponding to 5,000-50,000 photons per second at 1 micron entering the spectrograph calibration fiber. As typical astronomical exposures are on order of minutes, we expect the comb will have no difficulty achieving adequate signal-to-noise in our modal noise mitigation system during typical science exposures. Previous on-sky uses of a NIR \lq{}astro-comb\rq{} required a series of neutral density filters to prevent saturation during short exposures, even when coupling though the very same integrating sphere into a lossy testbed spectrograph \citep{ycas12}

Exceptions, perhaps, are the U-Ne or Th-Ar lamps. If  efficiency  is indeed an issue then the diffuser can be used alone, with a converging lens focusing on the multimode fiber. This configuration is significantly less lossy than using an integrating sphere (we achieve a total efficiency of $2\times10^{-3}$) with modal noise mitigation exceeding the mechanical agitation case, but inferior to using both the dynamic diffuser and integrating sphere, though it may be sufficient for many applications.

\section{Radial Velocities \& Modal Noise}
 Modal noise severely limited our RV precision during tests with a NIR laser comb calibrator \citep{ycas12, redman12}, and we {\bf must} eliminate this noise source as we develop the new generation of stabilized NIR spectrographs.  The changing speckle pattern on the fiber output manifests itself as a change in the centroid position (an RV change) of the line  since each speckle maps to a slightly different wavelength position in the dispersion direction.
 
  While we have shown that the dynamical diffuser significantly reduces the higher spatial frequencies in an image of the fiber output, it is instructive to perform an experiment to explicitly map the modal noise to achievable velocity precision.

We simulate the RV impact by acquiring a number of  three second exposures  using the 1550 nm laser, the Optotune diffuser, integrating sphere and an un-agitated 200 $\mu$m fiber coupled to the integrating sphere. Image sets were acquired with the diffuser in the static and dynamic mode.  Between each set of exposures the fiber bend radius and positions were changed using our mechanical scrambler, and the resulting mode pattern allowed to settle. No mechanical agitation was present during the exposures themselves. 

\begin{figure}
\begin{center}
\includegraphics[scale=.7]{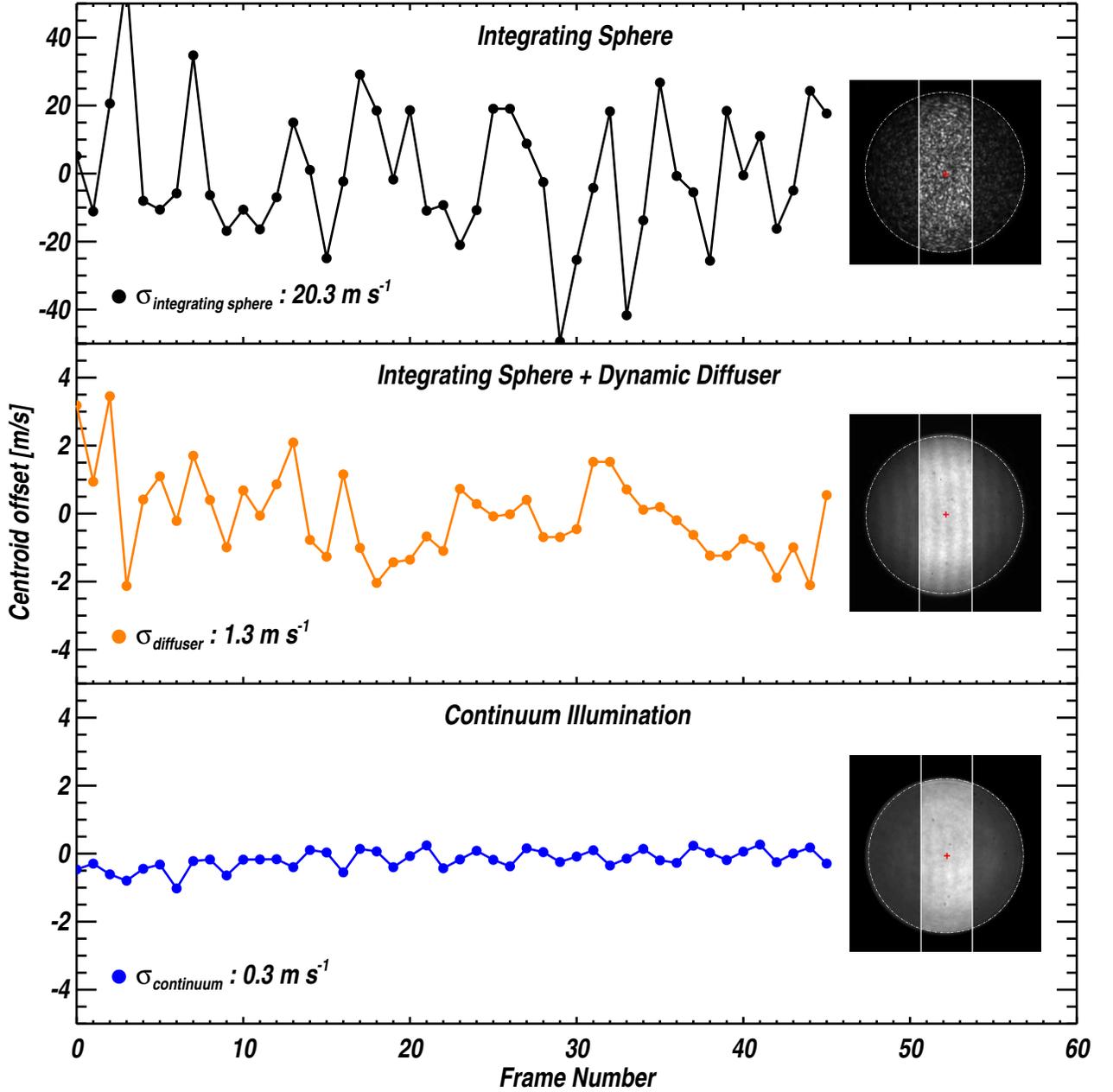}
\caption{Measured radial velocity scatter for images acquired with the integrating sphere (top), integrating sphere and dynamic diffuser (middle), and continuum-illuminated fiber (bottom). Image insets shows the speckle pattern recorded, and the region of the image that used to calculate the centroid and RV shifts. The 1.3 m s$^{-1}$ RV error  corresponds to the noise on {\bf a single emission line} on the spectrograph focal plane. Between each recorded image the fiber position was altered and the mode distribution allowed to settle. \label{rverr}}
\end{center}
\end{figure}

We use these images to simulate the case of the HPF spectrograph baseline \citep{mahadevan12}, a 300 $\mu$m fiber fed at f/3.65 with a  100 $\mu$m slit at spectrograph input. The 200 $\mu$m fiber fed at NA=0.22, has roughly the same number of modes as a 300 $\mu$m fiber fed at f/3.65. We impose a scaled digital slit on each of the recorded images and calculate the change in the centroid position in the dispersion direction. We use the expected resolution of the HPF spectrograph (R$\sim$ 50,000) to translate this centroid shift into a velocity shift. Figure \ref{rverr} shows the digital slit used for these calculations, as well as the resulting RV scatter for the case of the static and dynamic diffuser.  The results corroborate our experience \citep{ycas12, redman12} and show that modal noise is indeed a serious problem, leading to an RV scatter of ~20 m s$^{-1}$ even with the use of an integrating sphere and the static diffuser.  The dynamic diffuser significantly reduces this, yielding an RV scatter of only 1.3 m s$^{-1}$.  To probe the measurement precision of our setup in the absence of modal noise we replace the input SMF with a multimode fiber illuminated by a continuum source. Figure \ref{rverr} (bottom) shows that our experimental precision is 0.3 m s$^{-1}$. The 1.3 m s$^{-1}$ measurement is likely subject to other systematics (detector inhomogenities, pixel QE variability, flat fielding, etc) as well as the stability of the experimental setup and not necessarily the performance floor for the diffuser. This measurement also corresponds to the centroid precision {\bf of only one line on the spectrograph focal plane}. Assuming 30 -- 50 such emission lines across a typical echelle order yields a modal noise limited wavelength solution error of less than 15-20 cm s$^{-1}$. The use of the dynamical diffuser will therefore enable high precision RVs to be acquired even with narrow emission line sources.

\section{Discussion}

We have demonstrated a commercial off-the-shelf  solution to the fiber modal noise problem that has previously limited the use of narrow emission line calibration sources with high resolution astronomical spectrographs. Our solution uses a  dynamic diffuser \citep{blum12} and integrating sphere to randomize the modes. The dynamical diffuser changes the mode distribution at a timescale much shorter than the typical spectrograph exposure, and the integrating sphere further randomizes this distribution.   Our solution is also quite general in that any combination of a dynamic mode re-distributor and a mode mixer will help mitigate modal noise, though not necessarily as well as the setup we present. For example a galvano scanner may be used as a replacement to the dynamic diffuser in conjuction with an integrating sphere. Experiments coupling light from the diffuser directly to a multimode fiber achieve a total light throughput of $\sim$0.2\%, with modal noise mitigation exceeding the mechanical agitation case. Such solutions are useful with calibration sources like Th-Ar and U-Ne where minimizing light loss becomes important. 

The commercial Optotune diffuser  is made of polycarbonate, and has shallow absorption bands at certain wavelengths ($\sim$ 1180nm \& 1400nm)\footnote{Optotune Data Sheet: http://www.optotune.com/images/products/Optotune\%20LSR-3000\%20Series.pdf}.  It is possible to create custom engineered diffusers made of glass or fused silica \citep{sales06} that do not exhibit such absorption. Light weight diffusers can be agitated with the electroactive polymers, while heavier diffusers can utilize a rotating mount, albeit at lower frequencies. 

While light loss with the use of a integrating sphere makes it impractical to use with starlight, modal noise is less problematic in this case due to lower  visibility ($\nu$). Some modal mixing may be accomplished with custom diffusers between two fibers, but requires very small diffusion angles to minimize light loss and focal ratio degradation.  While exploration of such techniques may be fruitful, our opinion is that the star fiber modal noise is likely best addressed with a small amount of gentle agitation \citep{baud01,plavchan}, while a solution to the calibration fiber modal noise problem is presented in this paper for all optical and NIR wavelengths.

\acknowledgements
We acknowledge support from NSF  grants AST-0906034, AST-1006676,  AST-1126413, AST-1310885, the NASA Astrobiology Institute (NNA09DA76A), and the Center for Exoplanets and Habitable Worlds.


\clearpage

\begin{thebibliography}{}
\bibitem[Baudrand 
\& Walker(2001)]{baud01} Baudrand, J., \& Walker, G.~A.~H.\ 2001, \pasp, 113, 851

\bibitem[Blum et al.(2012)]{blum12} Blum, M., B{\"u}eler, M., 
Gr{\"a}tzel, C., Giger, J., \& Aschwanden, M.\ 2012, \procspie, 8252,  

\bibitem[Bland-Hawthorn 
\& Kern(2009)]{2009OExpr..17.1880B} Bland-Hawthorn, J., \& Kern, P.\ 2009, Optics Express, 17, 1880 


\bibitem[Bouchy et 
al.(2013)]{bouchy13} Bouchy, F., D{\'{\i}}az, R.~F., H{\'e}brard, G., et al.\ 2013, \aap, 549, A49 

\bibitem[Braje et al.(2008)]{braje08} Braje, D.~A., Kirchner, 
M.~S., Osterman, S., Fortier, T., 
\& Diddams, S.~A.\ 2008, European Physical Journal D, 48, 57


\bibitem[Dumusque et al.(2012)]{dumusque12} Dumusque, X., Pepe, 
F., Lovis, C., et al.\ 2012, \nat, 491, 207 

\bibitem[Goodman 
\& Rawson(1981)]{gr81} Goodman, J.~W., \& Rawson, E.~G.\ 1981, Optics Letters, 6, 324 
\bibitem[Grupp(2003)]{grupp03} Grupp, F.\ 2003, \aap, 412, 897 

\bibitem[Halverson et al.(2012)]{halverson12} Halverson, S., 
Mahadevan, S., Ramsey, L., et al.\ 2012, arXiv:1209.2704 

\bibitem[Hunter 
\& Ramsey(1992)]{hr92} Hunter, T.~R., \& Ramsey, L.~W.\ 1992, \pasp, 104, 1244 

\bibitem[Lemke et al.(2011)]{lemke11} Lemke, U., Corbett, J., 
Allington-Smith, J., \& Murray, G.\ 2011, \mnras, 417, 689 

\bibitem[Lo Curto et al.(2012)]{locurto12} Lo Curto, G., 
Manescau, A., Avila, G., et al.\ 2012, \procspie, 8446,  

\bibitem[Mahadevan et al.(2012)]{mahadevan12} Mahadevan, S., 
Ramsey, L., Bender, C., et al.\ 2012, \procspie, 8446,  

\bibitem[McCoy et al.(2012)]{mccoy12} McCoy, K.~S., Ramsey, L., 
Mahadevan, S., Halverson, S., \& Redman, S.~L.\ 2012, \procspie, 8446,

\bibitem[M{\'e}gevand et al.(2012)]{megevand12} M{\'e}gevand, D., 
Zerbi, F.~M., Cabral, A., et al.\ 2012, \procspie, 8446,

\bibitem[Murphy et al.(2007)]{murphy07} Murphy, M.~T., Udem, T., 
Holzwarth, R., et al.\ 2007, \mnras, 380, 839 

\bibitem[Osterman et al.(2012)]{osterman12} Osterman, S., Ycas, 
G.~G., Diddams, S.~A., et al.\ 2012, \procspie, 8450,

\bibitem[Phillips et al.(2012a)]{phillips12a} Phillips, D.~F., 
Glenday, A.~G., Li, C.-H., et al.\ 2012, Optics Express, 20, 13711 

\bibitem[Phillips et al.(2012b)]{phillips12b} Phillips, D.~F., 
Glenday, A., Li, C.-H., et al.\ 2012, \procspie, 8446, 

\bibitem[Plavchan et al.(2013)]{plavchan} Plavchan, P.~P., 
Bottom, M., Gao, P., et al.\ 2013, \procspie, 8864, 

\bibitem[Quinlan et al.(2010)]{quinlan10} Quinlan, F., Ycas, G., 
Osterman, S., 
\& Diddams, S.~A.\ 2010, Review of Scientific Instruments, 81, 063105 

\bibitem[Quirrenbach et al.(2012)]{quirrenbach12} Quirrenbach, A., 
Amado, P.~J., Seifert, W., et al.\ 2012, \procspie, 8446,  

\bibitem[Rawson et al.(1980)]{rawson80} Rawson, E.~G., Norton, 
R.~E., 
\& Goodman, J.~W.\ 1980, Journal of the Optical Society of America (1917-1983), 70, 968 

\bibitem[Redman et al.(2011)]{redman11} Redman, S.~L., Lawler, 
J.~E., Nave, G., Ramsey, L.~W., \& Mahadevan, S.\ 2011, \apjs, 195, 24 

\bibitem[Redman et al.(2012)]{redman12} Redman, S.~L., Ycas, 
G.~G., Terrien, R., et al.\ 2012, \apjs, 199, 2 

\bibitem[Sales et al.(2006)]{sales06} Sales, T.~R.~M., 
Schertler, D.~J., \& Chakmakjian, S.\ 2006, \procspie, 6290,  


\bibitem[Snellen et al.(2013)]{snellen13} Snellen, I.~A.~G., de 
Kok, R.~J., le Poole, R., Brogi, M., \& Birkby, J.\ 2013, \apj, 764, 182 

\bibitem[Wildi et al.(2012)]{wildi12} Wildi, F., Chazelas, B., 
\& Pepe, F.\ 2012, \procspie, 8446, 

\bibitem[Ycas et al.(2012a)]{ycas12} Ycas, G.~G., Quinlan, F., 
Diddams, S.~A., et al.\ 2012, Optics Express, 20, 6631

\bibitem[Ycas et al.(2012b)]{ycas12b} Ycas, G., Osterman, S., 
\& Diddams, S.~A.\ 2012, Optics Letters, 37, 2199


\end{thebibliography}
\end{document}